\documentclass[runningheads]{llncs}

\usepackage[T1]{fontenc}
\usepackage{graphicx}
\usepackage{epsfig}
\usepackage{calc}
\usepackage{amssymb,amsfonts}
\usepackage{textcomp}
\usepackage{xcolor}
\usepackage{comment}
\usepackage{appendix}
\usepackage{float}

\usepackage{subfig}
\newcommand{\beq}{\begin{equation}}
\newcommand{\eeq}{\end{equation}}
\usepackage{amstext}
\usepackage{amsmath}
\usepackage{amsthm}
\usepackage{multicol}
\usepackage{pslatex}
\begin{document}
\title{Complexity of Activity Patterns in a Bio-Inspired Hopfield-Type Network in Different Topologies}
%
%\titlerunning{Abbreviated paper title}
% If the paper title is too long for the running head, you can set
% an abbreviated paper title here
%
\author{Marco Cafiso \inst{1, 2}\orcidID{0009-0004-9519-1221} \and
Paolo Paradisi \inst{2,3}\orcidID{0000-0002-1036-4583}}
\authorrunning{M. Cafiso and P. Paradisi}
% First names are abbreviated in the running head.
% If there are more than two authors, 'et al.' is used.
%
\institute{Department of Physics 'E. Fermi', University of Pisa, Largo Bruno Pontecorvo 3, I-56127, Pisa, Italy \and
Institute of Information Science and Technologies ‘A. Faedo’, ISTI-CNR, Via G. Moruzzi 1, I-56124, Pisa, Italy \and
BCAM-Basque Center for Applied Mathematics, Alameda de Mazarredo 14, E-48009, Bilbao, BASQUE COUNTRY, Spain \\
\email{marco.cafiso@phd.unipi.it, paolo.paradisi@cnr.it}}

\maketitle              % typeset the header of the contribution
\begin{abstract}
Neural network models capable of storing memory have been extensively studied in computer science and computational neuroscience. The Hopfield network is a prototypical example of a model designed for associative, or content-addressable, memory. This model has been widely studied, and various versions have been proposed.
%Biological neural networks and their complex features have been investigated over the last twenty years. The complex topological aspects were also recently investigated in artificial neural networks and learning. 
Further, ideas and methods from complex network theory have been incorporated into artificial neural networks and learning, emphasizing their structural properties. Nevertheless, the temporal dynamics also play a vital role in biological neural networks, whose temporal structure is a crucial feature to examine.
 %in systems that display complex intermittency, such as biological neural networks. 
Biological neural networks 
%belong to a particular class of dynamical systems that 
display complex intermittency and, thus, can be studied through the lens of the temporal complexity (TC) theory. The TC approach 
look at the metastability of self-organized states, characterized by a power-law decay in the inter-event time distribution and in the total activity distribution or a scaling behavior in the corresponding event-driven diffusion processes.
In this study, we present a temporal complexity (TC) analysis of a biologically-inspired Hopfield-type neural network model. We conducted a comparative assessment between scale-free and random network topologies, with particular emphasis on their global activation patterns.
Our parametric analysis revealed comparable dynamical behaviors across both neural network architectures. Furthermore, our investigation into temporal complexity characteristics uncovered that seemingly distinct dynamical patterns exhibit similar temporal complexity behaviors. In particular, similar power-law decay in the activity distribution and similar complexity levels are observed in both topologies, but with a much reduced noise in the scale-free topology.
Notably, most of the complex dynamical profiles were consistently observed in scale-free network configurations, thus confirming the crucial role of hubs in neural network dynamics.
\end{abstract}
\section{Introduction}
\label{sec:intro}

The Hopfield model of associative memory %\cite{hopfield_pnas1982} 
is the first example of recurrent neural network (RNN) and is widely studied in various scientific disciplines. Hopfield initially proposed this network paradigm in \cite{hopfield_pnas1982} and elaborated its theoretical foundations in subsequent works \cite{hopfield_pnas1984,hopfield_n1995}. 
Moreover, he developed an enhancement to his original model, adapting it to a continuous-time leaky-integrate-and-fire neuronal system \cite{hopfield_pnas1984}, additionally exploring neurons with graduated responses—namely, those utilizing a sigmoid function to process voltage inputs from connected neurons. The fundamental property of the Hopfield model lies in its memory storage process which is based on the Hebbian principle \cite{hebb_1949}, commonly condensed into the phrase: "neurons that activate simultaneously strengthen their connections" \cite{lowel_s1992}. Its importance is due to the associative recall process characterized by energy minimization dynamics, where the system naturally moves toward lower energy states through a descent mechanism. This gradual reduction in energy corresponds to the convergence toward stored memory patterns, allowing the system to retrieve complete information from partial inputs by following the downward slope of the energy landscape until reaching a stable minimum. The Hopfield neural model and its variants properly belong to the category of Spiking Neural Networks (SNNs). These biologically-inspired networks continue to generate significant research interest due to their remarkable energy efficiency and sophisticated temporal information processing capabilities. Although current implementations of SNNs have not yet achieved the performance levels of conventional deep neural networks, they exhibit extraordinary promise in the emerging field of neuromorphic computing \cite{davies_ieeem2018}. 

% \noindent
% An important behavior highlighted by Hopfield himself is the manifestation of self-organizing behavior concerning memory stability.
\noindent
An important behavior highlighted by Hopfield himself is the manifestation of cooperative behavior with regard to the stabilization of stored patterns.
%in the network model. 
In this context, Grinstein and Linsker %in 2005 
\cite{grinstein2005model} explored how network architecture influences a neural system that builds upon the Hopfield model, enhancing its biological realism while partially preserving the computational benefits of binary McCulloch-Pitts neurons compared to fully continuous-time implementations. 
In particular, their proposal incorporated both maximum firing duration and refractory periods into individual neuronal dynamics. 

\noindent
The study of topological complexity in neural networks is now crucial across different scientific domains, particularly in computational neuroscience and artificial intelligence (see, e.g., \cite{kaviani_eswa2021} for a survey). The structural organization of these networks significantly influences their computational capabilities, learning efficiency, and information processing dynamics. Understanding how network topology shapes emergent properties has become essential for advancing both our knowledge of biological neural systems and the development of more efficient artificial neural architectures. Indeed, different works highlighted how different complex network topologies, such as random (Erd$\ddot o$s-R\' enyi) \cite{erdos1959random}, \cite{gros_2013}, small-world or scale-free networks \cite{boccaletti_pr2006}, outperform artificial neural networks with all-to-all connectivity \cite{adjodah_proceed2020,kaviani_icte2020,kaviani_eswa2021,lu_pla2006,mcgraw_pre2003,shafiee_ieee-a2016,torres_nc2004}.

\noindent
The term \textit{complexity} refers to the ability of a multi-component system to exhibit self-organizing behavior, a characteristic demonstrated through the formation of coherent states across both space and time \cite{grigolini_csf15_bio_temp_complex,paradisi_springer2017,paradisi_csf15_preface}. In many scientific disciplines, however, complexity is narrowly associated with spatial patterns alone – a perspective inherited from graph theory and complex network frameworks \cite{albert_rmp2002,barabasi_s1999,barabasi_nrg2004,watts_n1998} - and the temporal aspects are often overlooked. Nevertheless, temporal behaviors constitute another vital aspect of complex self-organizing systems, which are not only influenced by topological configurations but also by the internal dynamic properties of each network component at various scales—ranging from individual nodes, to clusters of nodes, and ultimately to the entire network. In this framework, Temporal Complexity (TC) – also termed Intermittency-Driven Complexity (IDC) – describes a system’s capacity to produce metastable self-organized states in its temporal dynamics \cite{grigolini_csf15_bio_temp_complex,paradisi_springer2017,grigolini_pre11}. These states exhibit transient durations punctuated by abrupt transitions between distinct dynamical regimes. In its basic formulation, the TC/IDC theoretical framework is inspired to the Cox's renewal theory \cite{cox_1970_renewal}. In this context, Cox’s failure processes are reconceptualized through a temporal lens – specifically recast as abrupt state transitions or discontinuous jumps in the system’s observable dynamics. These rapid transition events mark the passages between two self-organized states or between a self-organized state and a disordered or non-coherent state. The TC look at the series of transition events as a point process to determine if the renewal condition\footnote{
%%%%%%%%%
Renewal condition is defined by the mutual independence of successive events and, thus, of successive inter-event times.
%%%%%%%%%
} is verified \cite{bianco_cpl07,cox_1970_renewal,paradisi_cejp09}, that is not easy to assess due to the mixed secondary, or spurious, events in the series \cite{paradisi_csf15_pandora,paradisi_springer2017}. Different examples of self-organizing behaviors can be found in nature, e.g., eddies in a turbulent flow or synchronization epochs
in neural dynamics, and the identification of events is usually achieved employing 
proper event detection algorithm in signal processing \cite{paradisi_springer2017,paradisi_chapter2023}. 

\noindent
The manifestation of complex self-organized states is associated with power-law scaling behaviors. These scaling behaviors can be observed in the probability distributions of some physical observables, such as inter-event time series and neural avalanche sizes \cite{beggsplenz_jn2003}. The fundamental aspect in the TC theory relies on the evaluation of the inter-event times (IET) probability density function (PDF), whose pattern essentially define the TC features.
However, as said above, the direct evaluation of TC indices from the IET-PDF is often untrustworthy due to the noisy secondary events in the series or other side effects \cite{allegrini_pre10,paradisi_csf15_pandora}.
%can be difficult 
When noisy
events are mixed with genuine complex events, i.e., events associated with self-organizing mechanisms,
a reliable technique used to assess the IET-PDF that involves also several well-known scaling analyses is the Event-Driven Diffusion Scaling (EDDiS) algorithm \cite{paradisi_csf15_pandora,paradisi_springer2017}. This approach is based on the finding that, under wide assumptions, event-driven diffusion processes can enhance the complex process with respect to the superposed noisy process. This is obtained by looking at the scaling that the diffusion process inherits from the driver, i.e., the complex event sequence, even when it is blurred by secondary events.
The EDDiS algorithm has also been effectively utilized in the analysis of brain data, demonstrating its ability to distinguish various brain states, ranging from wakefulness and relaxed conditions to different sleep stages \cite{allegrini_csf13,allegrini_pre15,paradisi_aipcp13}. 

% Moreover, a recent study highlights some correlations between the topology of a very simple Spiking Neural Network (SNN) and the EDDiS scaling indexes in some specific dynamical cases \cite{cafiso_2024}.

\noindent
Currently, the connections between a network's temporal dynamics and its structural complexity remain poorly understood, as does the relationship between connectivity patterns and learning capabilities. A recent study highlights some correlations between the topology of a very simple Spiking Neural Network (SNN) and the EDDiS scaling indexes in some specific dynamical cases \cite{cafiso_2024}.
This study presents findings on the temporal-structural relationship, with an eye toward future applications in optimizing network connectivity for enhanced learning. We conduct TC analysis on biologically-inspired Hopfield-type neural networks, comparing scale-free versus random network architectures.
Section \ref{sec:model} outlines our bio-inspired Hopfield-type neural model and details our methodology for generating different network topologies. Section \ref{sec:results} presents our numerical simulation outcomes and corresponding TC analyses, which are then interpreted in Section \ref{sec:discussion}. We conclude with key insights and future research directions in Section \ref{sec:conclusion}. Finally, in Appendix \ref{appendix:eddis} we provide a concise overview on the event-driven diffusion scaling analysis (EDDiS) used to explore the TC features of these networks.

%%%%%%%%%%%%%%%%%%%%%%%%%%%%%%%%%%%%%%%%%%%%%%%%%%%%%%%
%%%%%%%%%%%%%%%%%%%%%%%%%%%%%%%%%%%%%%%%%%%%%%%%%%%%%%%
\section{Model description}
\label{sec:model}

%%%%%%%%%%%%%%%%%%%%%%%%%%%%%%%%%%%%%%%%%%%%%%%%%%%%%%%%%%%
\subsection{Bio-Inspired Hopfield-type network model}

In this work of 2005 \cite{grinstein2005model}, Grinstein and Linsker modify the classical Hopfield network dynamics by adding three bio-inspired elements: (\(i\)) a random endogenous firing probability \(p_{endo}\) for each node; (\(ii\)) a maximum firing duration, thanks to which the activity of a node shuts down after \(t_{max}\) consecutive time steps; (\(iii\)) a refractory period such that a node, once activated and subsequently deactivated, must remain inactive for at least \(t_{ref}\) consecutive time-steps. 
In this model, each neuron $i$ has two states: \(S_i = 0\) ("not firing") and \(S_i = 1\) ("firing at maximum rate"). The weight of link from $j$ to $i$ is
given by \(J_{ij}\) (Non-connected neurons have \(J_{ij} = 0\)). 
The network is initialized at time \(t = 0\) by randomly setting each neuron state \(S_i(0)\) equal to 1 with a probability \(p_{init}\) that we chose equal to the endogenous firing probability (\(p_{init} = p_{endo}\)). At each time step the weighted input to node \(i\) is defined:
\begin{equation} \label{grinstein_weigthed_input_to_node_j}
    I_i(t) = \sum_j J_{ij}S_j(t)
\end{equation}
as in the classical Hopfield network dynamics. At each time-step, the state of the node \(i\) evolves according to the following rules:
\begin{enumerate}
    \item If \(S_i(\tau)\) for all \(\tau = t, t-1,\cdots,t-t_{max}+1\), then \(S_i(t+1) = 0\) (maximum firing duration rule).
    \item If \(S_i(t) = 1\) and \(S_i(t+1) = 0\) then $S(\tau)=0$ for 
    \( (t + 1) < \tau \leq (t + t_{ref}) \)  
  %  for at least one \(\tau\) satisfying \((t - t_{ref}) \leq \tau \leq (t - 1)\), then \(S_i(t+1) = 0\) 
  (refractory period rule).
    \item If neither rule 1 nor rule 2 applies, then
    \begin{enumerate}
        \item If \(I_i(t) \geq b_i\) then \(S_i(t + 1) = 1\);
        \item If \(I_i(t) < b_i\) then \(S_i(t + 1) = 1\) with a probability equal to \(p_{endo}\) otherwise \( S_i(t + 1) = 0 \).
    \end{enumerate}
    where $b_i$ is the firing threshold of neuron $i$.
\end{enumerate}

In this study, we chose to maintain fixed the link's weights and the firing threshold across the network: \(J_{ij} = J\) and \( b_i = b \).

%%%%%%%%%%%%%%%%%%%%%%%%%%%%%%%%%%%%%%%%%%%%%%%%%%%%%%%%%%%
\subsection{Network topology: scale-free vs. random}

Here, we consider two network topologies: Scale-Free and Random (Erd$\ddot o$s-R\'enyi). Our networks are constrained to have the same minimum number $k_0$ of outgoing links for each neuron and the same average out-degree value\footnote[1]{
%%%%%%
The degree of a node in a network is the number of links of the node itself. In a directed network, each node has an out-degree, given by the number of outgoing links, and an in-degree, given by the number of incoming links.
%%%%%%%%
} $\langle k \rangle$. In both cases, self-loops and multiple directed edges from one node to another are excluded, following the methodology outlined in \cite{grinstein2005model}. 
The first class of network topology is the Scale-Free (SF), characterized by a power-law node out-degree distribution. 
In particular, the probability of a node \(i\) to
have \(k_i\) outgoing links is given by: 
\begin{eqnarray}
&&\forall i = 1, ..., N:\ \ P_{_{\rm SF}}(k_i) = \frac{m}{k_i^{\alpha}} \\
\label{sf_1}
\ \nonumber \\
&&m = \frac{ \alpha-1 }{ k_0^{(1-\alpha)} - (N-1)^{(1-\alpha)} } 
\label{sf_2}
\end{eqnarray}
For developing SF networks, we used a power-law exponent \(\alpha = 2.5\). In particular, the algorithm used to develop SF networks is the following:
\begin{enumerate}       
    \item
    For each node \(i\), choose the out-degree \(k_i\) 
    as the nearest integer of the real number defined by:    
    \beq
        \label{SF_out_degree}
        k_i = (((N-1)^{(1-\alpha)} - k_0^{(1-\alpha)})\xi_i + k_0^{(1-\alpha)})^{\frac{1}{1 - \alpha}}
    \eeq
    being $\xi$ a random number uniformly distributed in $[0,1]$. This formula is obtained by the cumulative function method. The drawn $k_i$ are within the range $[k_0 , N-1]$.\\
    \item 
    Given $k_i$ for each node \(i\), the target nodes are selected by drawing $k_i$ integer numbers $\{j^i_1,...,j^i_{k_i} \}$ uniformly distributed in the set 
    $\{1,...,i-1,i+1,...,N \}$.      
    \item 
    Finally, the adjacency or connectivity matrix is defined as: 
    \beq
    \label{adjacency_sf}
       A^{^{SF}}_{ij} = \left\{
       \begin{array}{ll} & 1 \quad {\rm if}\ \  
       j\in T^i = \{ j^i_1,...,j^i_{k_i} \}\\
       & 0 \quad {\rm otherwise}
       \end{array}
       \right.
    \eeq
 With this choice, the in-degree distribution results in a mono-modal distribution similar to a Gaussian distribution.
\end{enumerate}

\noindent
The second class is that of Erd$\ddot o$s-R\'enyi (ER) graphs, which are random graphs where each pair of distinct nodes is connected with a probability \(p\). In an ER network with $N$ nodes and without self-loops, the average degree is simply
given by: $\langle k \rangle_{_{\rm ER}} = p_{_{\rm ER}} \left(N-1\right)$. Then, from the
equality of degree averages: $\langle k \rangle_{_{\rm ER}} = \langle k \rangle_{_{\rm SF}}$
we derive:
\beq
p_{_{\rm ER}} = \frac{\langle k \rangle_{_{\rm SF}}}{N-1}
\label{p_random}
\eeq
The theoretical mean out-degree of the SF network is approximated by the following formula:
$$
\langle k \rangle \simeq m \frac{k_0^{2-\alpha} - (N-1)^{2-\alpha}}{\alpha-2} = $$
$$
= \frac{\alpha-1}{\alpha-2} 
\frac{k_0^{2-\alpha} - (N-1)^{2-\alpha}}{k_0^{1-\alpha} - (N-1)^{1-\alpha}}
$$
that is obtained by considering $k$ as a continuous random variable. Although different statistical samples drawn from $P_{_{\rm SF}}$ can have very different mean out-degrees due to the large variability of SF degree distribution.
Thus, we have chosen to numerically evaluate the mean out-degree associated with the sample drawn from $P_{_{\rm SF}}$ and to use this value instead of the theoretical one to define $P_{_{\rm SF}}$. The algorithm used to generate this kind of network topology is as follows:

\begin{enumerate}
\item
From the adjacency matrix $A^{^{SF}}_{ij}$ the actual mean out-degree $\langle k \rangle_{_{\rm SF}}$ is computed.
\item
For each couple of nodes $(i,j)$ with $j \ne i$ a random number $\xi_{i,j}$ is drawn from a uniform distribution in $[0,1]$.
\item 
Finally, the adjacency matrix is defined as: 
  \beq
        \label{adjacency_er}
           A^{^{ER}}_{ij} = \left\{
           \begin{array}{ll} & 1 \quad {\rm if}\ \ 
            \xi_{i,j} < p_{_{\rm ER}}  \\
           & 0 \quad {\rm otherwise}
           \end{array}
           \right.
        \eeq
where $p_{_{\rm ER}}$ is given by Eq. \eqref{p_random}.
\end{enumerate}

%\vspace{-1cm}

%%%%%%%%%%%%%%%%%%%%%%%%%%%%%%%%%%%%%%%%%%%%%%%%%%%%%%%%%%%%%%%%%%%%%%%%
\section{Numerical simulations and results}
\label{sec:results}

\subsection{Parametric Analysis}

\noindent
%At first, 
We performed a qualitative parametric analysis by looking at the shape of the total activity distribution $P(n_c|N_c = 1)$ (see Appendix \ref{app:neural_coinc}) 
and the average activity over time. We fixed \(t_{max}\) value of 3, while systematically varying the other parameters. In particular, \(J \) which was constrained to integer values ranging from 1 to 4, \(p_{endo}\) set to $0.001$, $0.01$ and $0.1$, \(k_0\) limited to integers spanning from 1 to 5, \(b \) with two options of either 2 or 3, and \(t_{ref}\) set at 0, 4, 6, 8 and 10.
%
%For dimensional reasons, 
The model's dynamics depend only on the adimensional parameter \(\pi = \frac{J}{b}\) that is represented in the resulting plots. 

\begin{figure}[]
\centering
\includegraphics[scale=0.3]{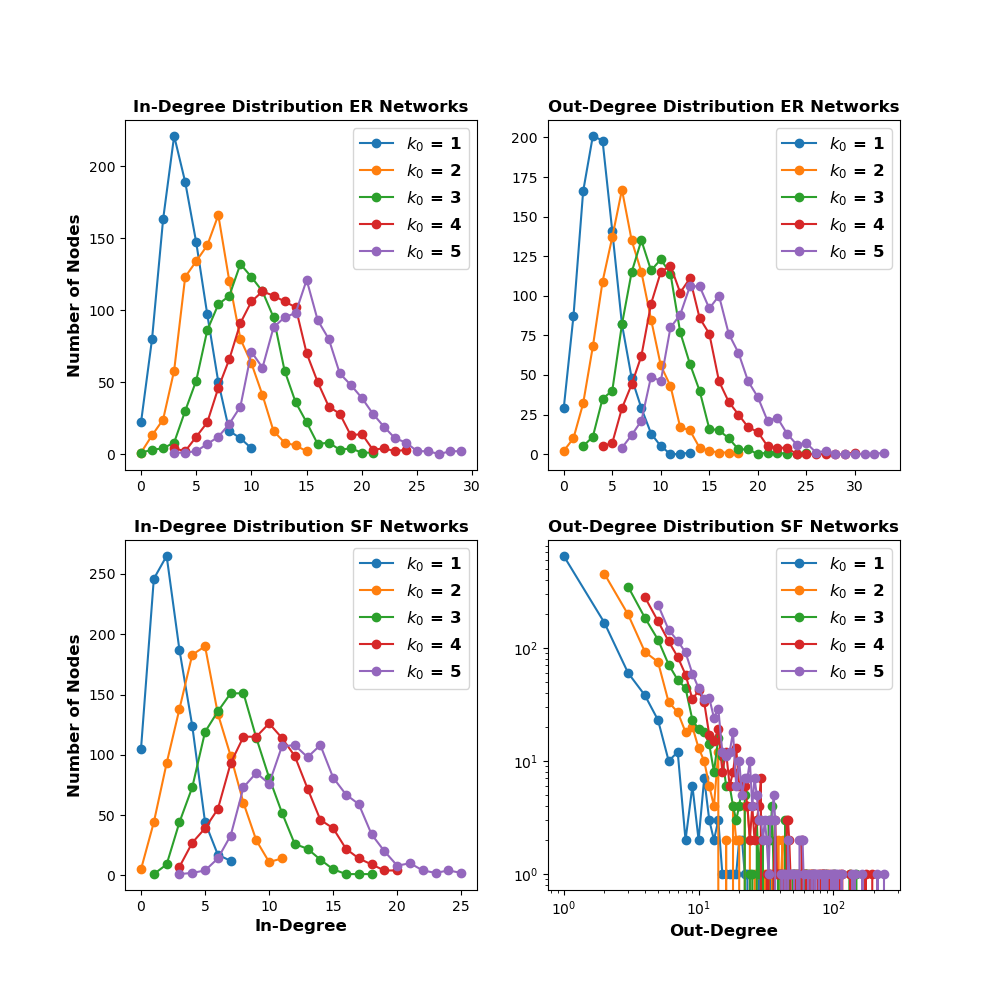}
\caption{In-degree and Out-degree distributions generated for Erd$\ddot o$s-R\'eny (ER) and Scale-Free (SF) network topologies with varying minimum out-degree values ($k_0$).} 
\label{fig:In_Out_Degrees}
\end{figure}

\noindent
Simulations were conducted over $20000$ time steps on networks containing $1000$ neurons. We generated five distinct topologies for each network configuration type by varying the minimum out-degree value ($k_0$). Table \ref{tab:total_number_links} presents the total number of links for each network variant, while Fig. \ref{fig:In_Out_Degrees} displays the corresponding in-degree and out-degree distributions.
%% Aggiungere parte su Degree Distributions (In- e Out-) e rispettivo numero di links (Fig. 1, Tab. 1)
%Fig. \ref{fig:In_Out_Degrees} presents the degree distributions for all developed networks. 
Notably, the in-degree distributions of the two network topologies exhibit no significant differences. Conversely, the out-degree distributions reveal the expected distinct characteristics: SF networks display a power-law behavior, while ER networks exhibit a slightly asymmetric mono-modal distribution similar to the in-degree distributions of both topologies, all approximating a Gaussian distribution as $k_0$ increases.
%Table \ref{tab:total_number_links} reports the total number of links for each generated network as a function of the minimum out-degree parameter ($k_0$). 
The results summarized in Table \ref{tab:total_number_links} indicate that
%while both SF and ER network configurations have a comparable total number of links, 
the ER configuration consistently contains more links than the SF configuration. Notably, while the number of links rapidly increases with $k_0$, the difference between ER and SF links remains almost constant. As a consequence, the difference between ER and SF connectivity results to be much more relevant for low $k_0$.
%%%%%%%%%%%%%%%%%%%%%%%%%%%%%%%%%%%%
%%%%%%%% TABLE TOT LINKS %%%%%%%%%%%
%%%%%%%%%%%%%%%%%%%%%%%%%%%%%%%%%%%%
%
% Table Total Links ER/SF
%
\begin{table}[h]
        \caption{\small Total Number of Links for Each Generated Network Across Different Minimum Out-degree Values ($k_0$). \normalsize}
        \begin{center}
        \begin{tabular}{c|c|c|c|c|c}
            % \cline{2-5}
            %  & \multicolumn{2}{|c|}{DFA ($H$)} & \multicolumn{2}{|c|}{DE ($\delta$)}\\
            % \cline{2-5}
            %  & Short-Time & Long-Time & Short-Time & Long-Time \\
            % \hline
            Min Links ($k_0$) & 
            1  & 
            2 & 
            3 & 
            4 &
            5\\
            \hline
            \hline
            ER Networks & 
            3760  & 
            6601 & 
            9343 & 
            12024 &
            14672\\
            \hline
            SF Networks & 
            2244  & 
            4833 & 
            7434 & 
            10036 &
            12647\\

        \end{tabular}
        \end{center}
    \label{tab:total_number_links}
\end{table}
%\vspace{-1cm}
%%%%%%%%%%%%%%%%%%%%%%%%%%%%%%%%%%%%%%%%
%%%%%%%%%% FIGURE DEGREE %%%%%%%%%%%%%%%
%%%%%%%%%%%%%%%%%%%%%%%%%%%%%%%%%%%%%%%%
%
% Degrees ER/SF
%
%
%\vspace{-1cm}
%

\begin{figure}[]
\centering
\includegraphics[scale=0.34]{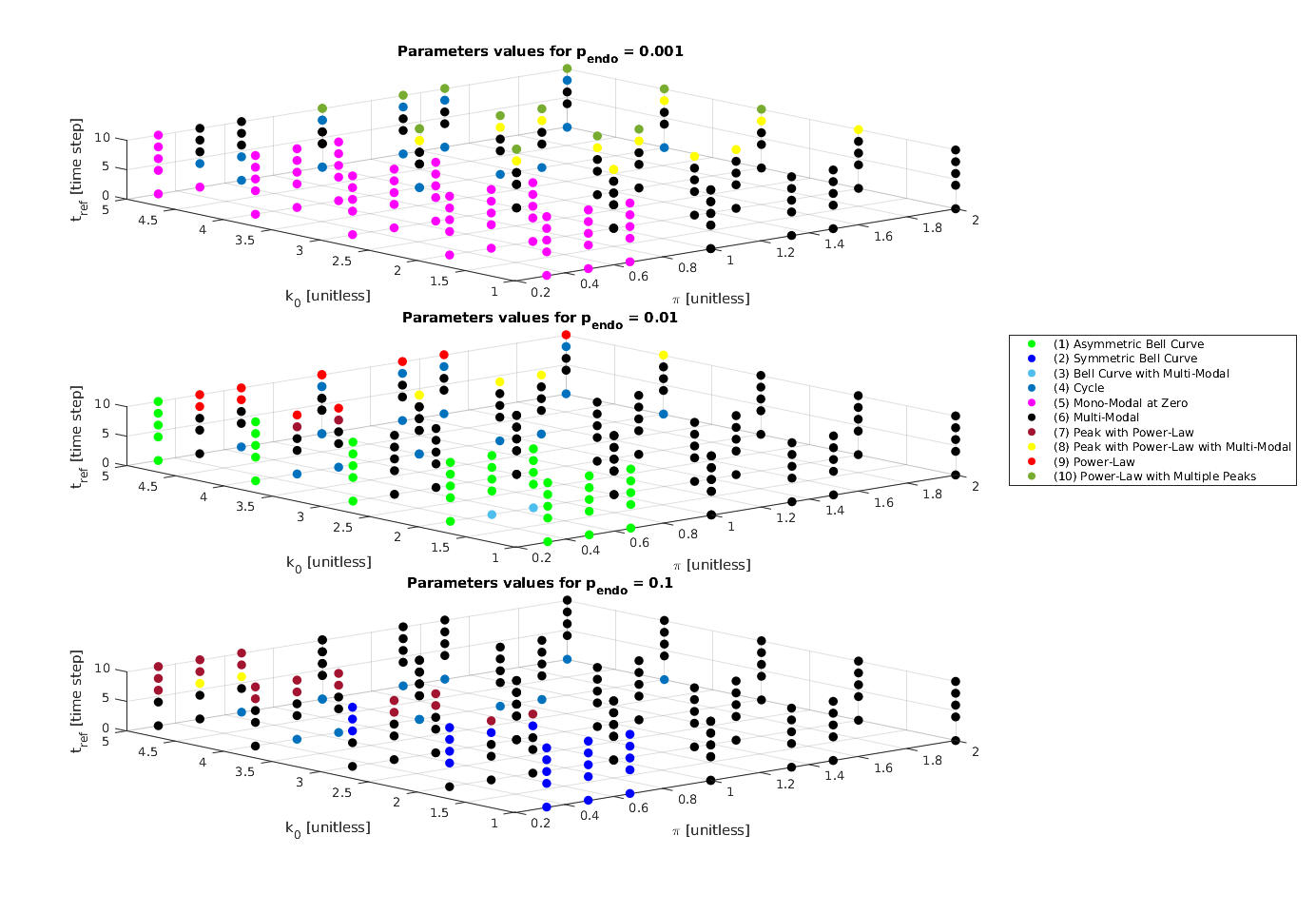}
\caption{Results of parameter analysis derived from the behavior of the total activity distribution in ER networks \cite{cafiso_2024}.} 
\label{fig:ER_parametric_analysis}
\end{figure}

\begin{figure}[]
\centering
\includegraphics[scale = 0.34]{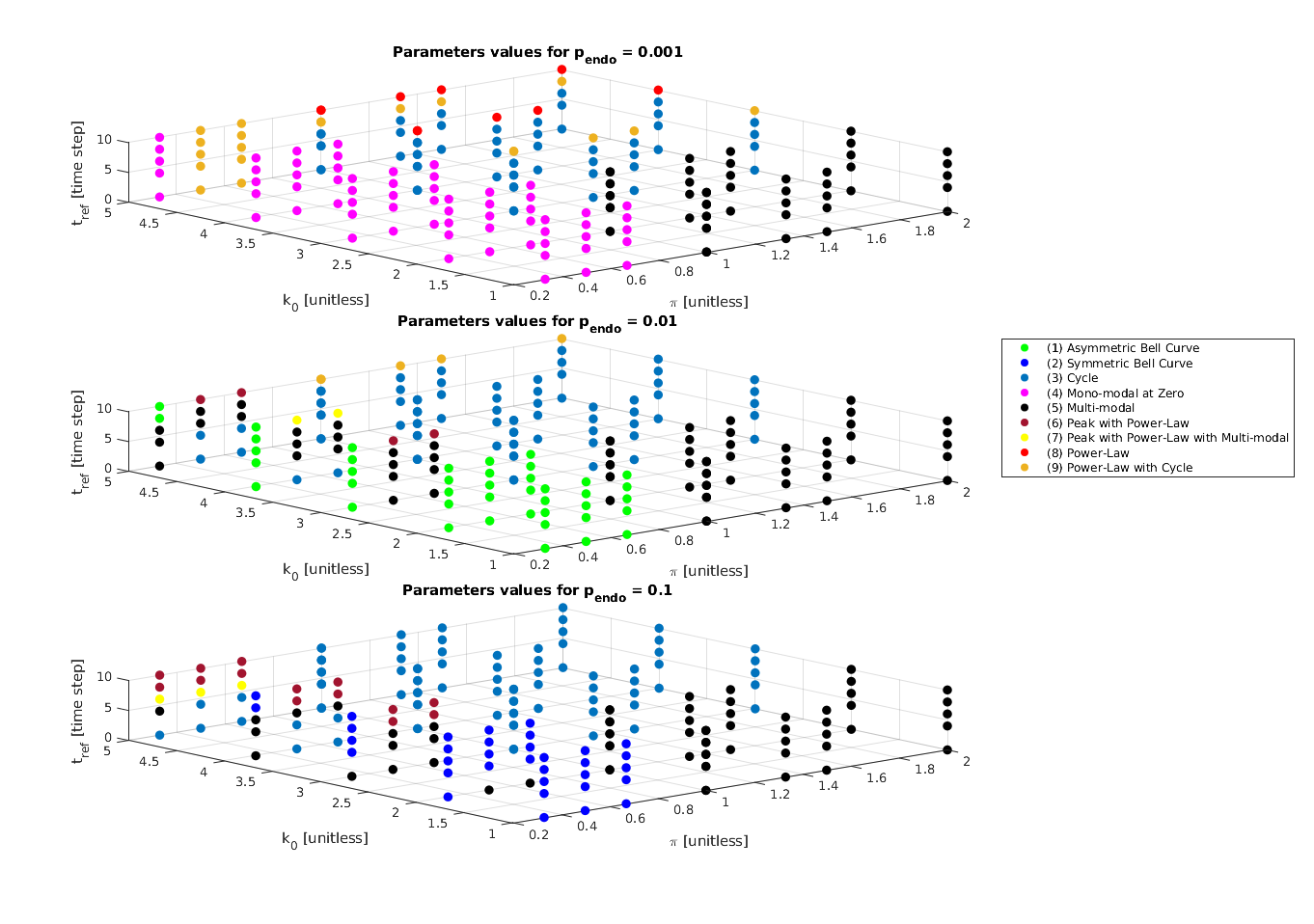}
\caption{Results of parameter analysis derived from the behavior of the total activity distribution in SF network \cite{cafiso_2024}.} 
\label{fig:SF_parametric_analysis}
\end{figure}

\begin{figure}[H]
\centering
\subfloat[Average Activity vs Time]{\includegraphics[scale=0.25]{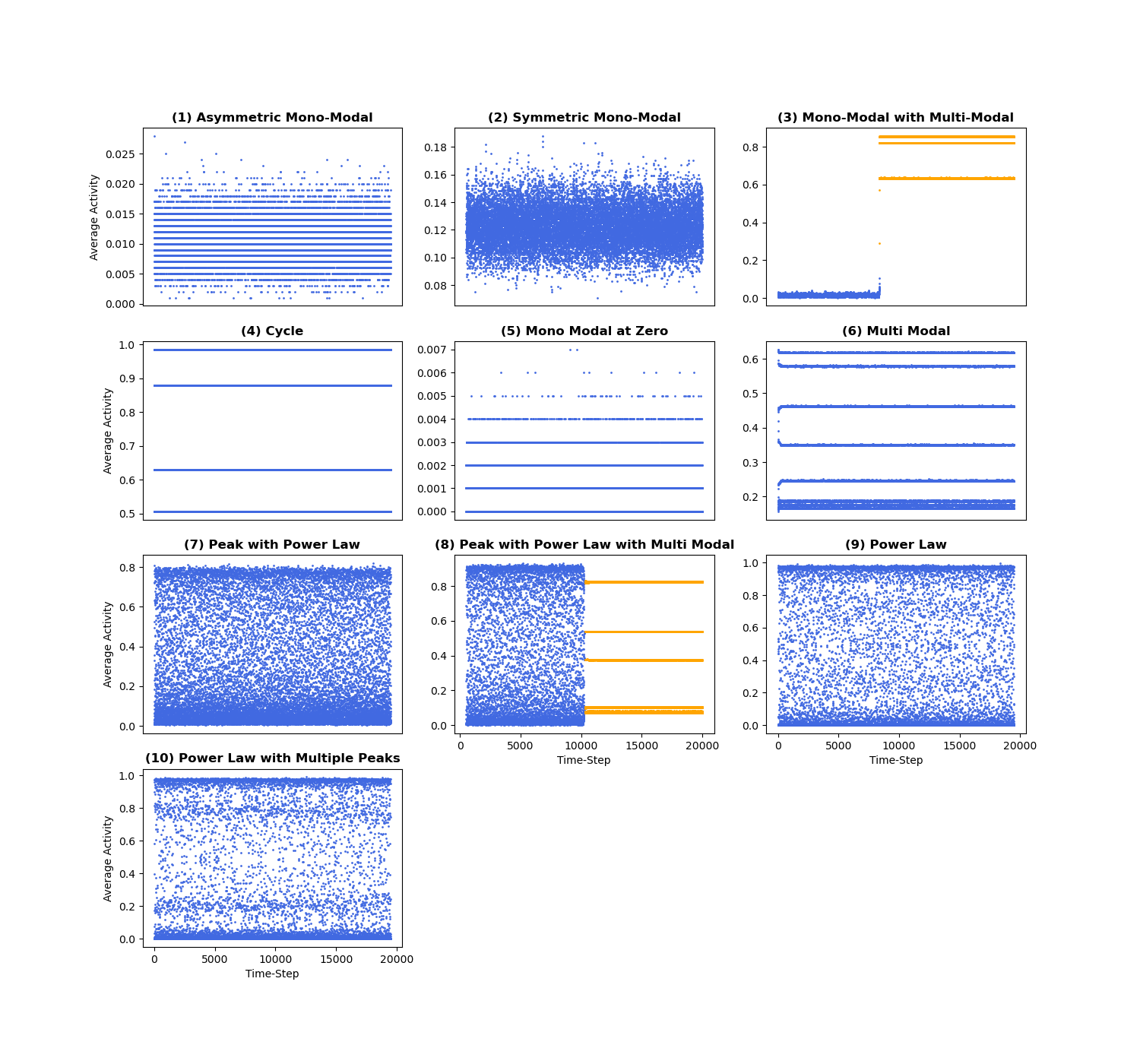}}
\hspace{-5.cm}
\subfloat[Histograms of Total Activity]{\includegraphics[scale=0.25]{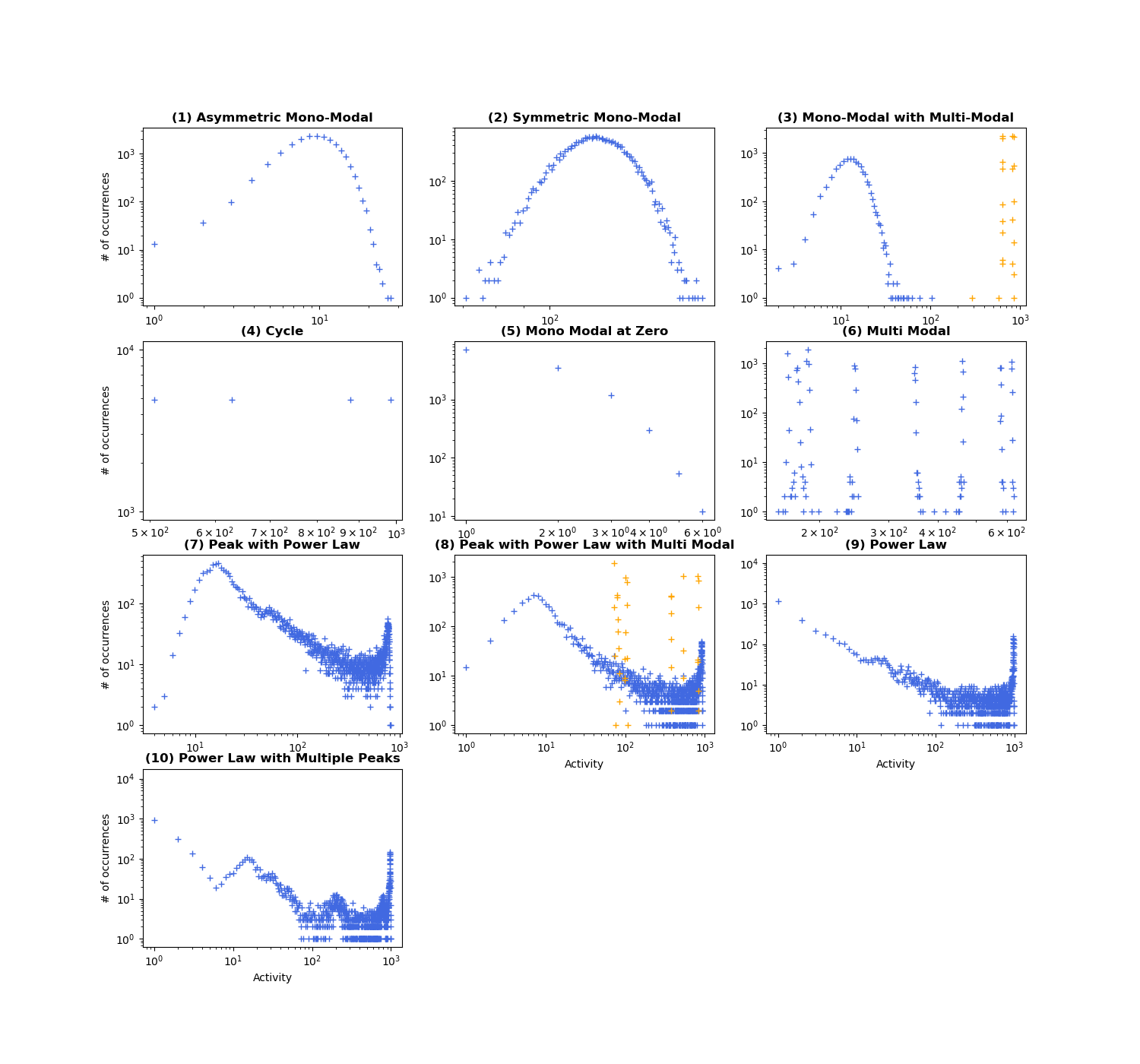}}
\caption{Qualitative analysis of ER network behaviors: (a) Average Activity plots over time and (b) Histograms of Total Activity 
%for all the qualitative behaviors found in ER networks. 
Parameters for each panel: (1) \(k_0\) = 1, \(t_{ref}\) = 4, \(b\) = 2, \(p_{endo}\) = 0.01, and \(J\) = 1; (2) \(k_0\) = 1, \(t_{ref}\) = 4, \(b\) = 2, \(p_{endo}\) = 0.1, and \(J\) = 1;  (3) \(k_0\) = 2, \(t_{ref}\) = 0, \(b\) = 2, \(p_{endo}\) = 0.01, and \(J\) = 1; (4) \(k_0\) = 4, \(t_{ref}\) = 0, \(b\) = 2, \(p_{endo}\) = 0.01, and \(J\) = 1; (5) \(k_0\) = 1, \(t_{ref}\) = 4, \(b\) = 2, \(p_{endo}\) = 0.001, and \(J\) = 1; (6) \(k_0\) = 3, \(t_{ref}\) = 6, \(b\) = 3, \(p_{endo}\) = 0.1, and \(J\) = 2; (7) \(k_0\) = 5, \(t_{ref}\) = 6, \(b\) = 3, \(p_{endo}\) = 0.1, and \(J\) = 1; (8) \(k_0\) = 5, \(t_{ref}\) = 6, \(b\) = 3, \(p_{endo}\) = 0.1, and \(J\) = 2; (9) \(k_0\) = 5, \(t_{ref}\) = 10, \(b\) = 2, \(p_{endo}\) = 0.01, and \(J\) = 3; (10) \(k_0\) = 5, \(t_{ref}\) = 10, \(b\) = 2, \(p_{endo}\) = 0.001, and \(J\) = 3 \cite{cafiso_2024}.}
\label{fig:ER_average_activity_total_activity}
\end{figure}

\begin{figure}[H]
\centering
\subfloat[Average Activity vs Time]{\includegraphics[scale=0.25]{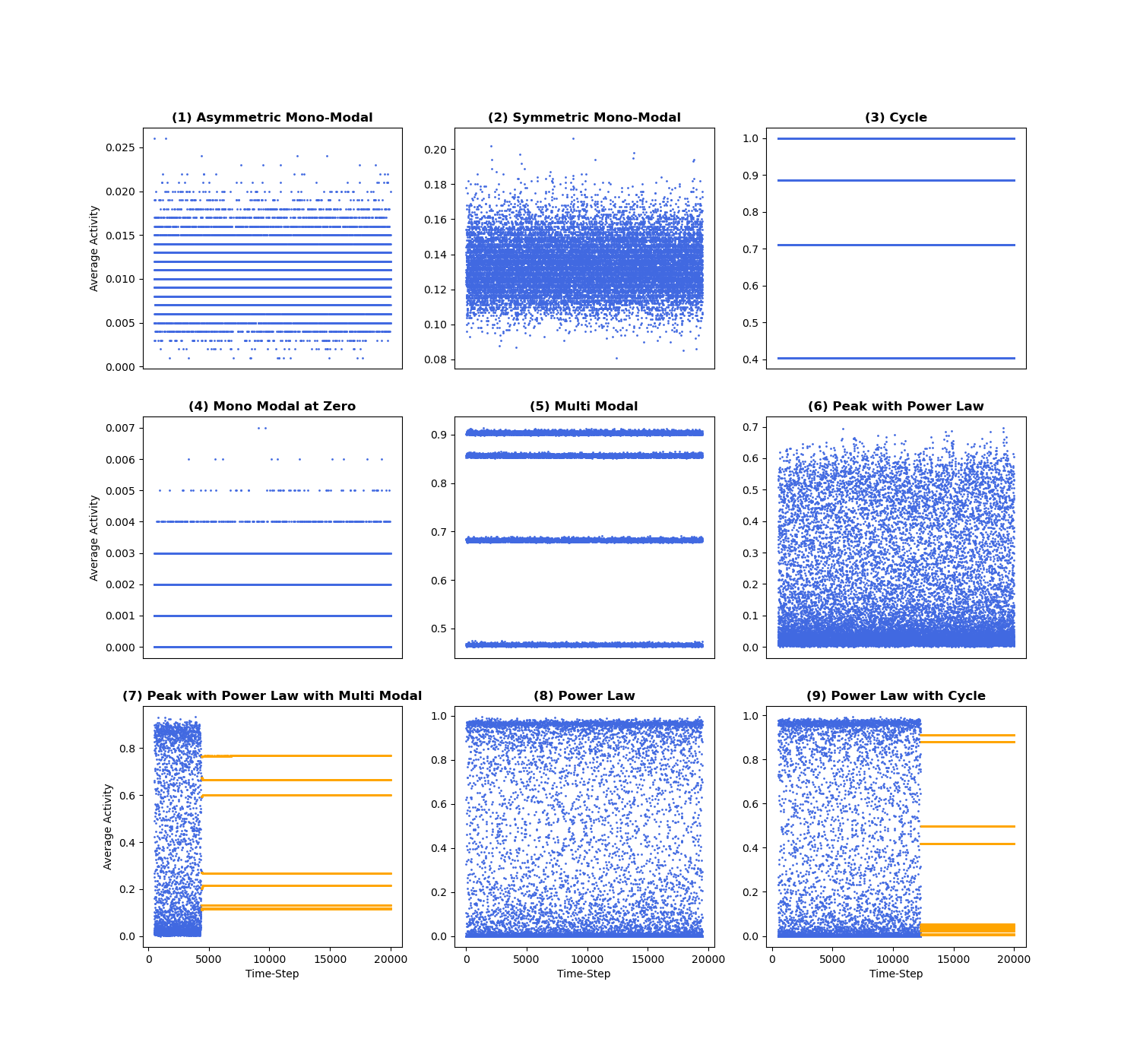}}
\hspace{5.cm}
\subfloat[Histograms of Total Activity]{\includegraphics[scale=0.25]{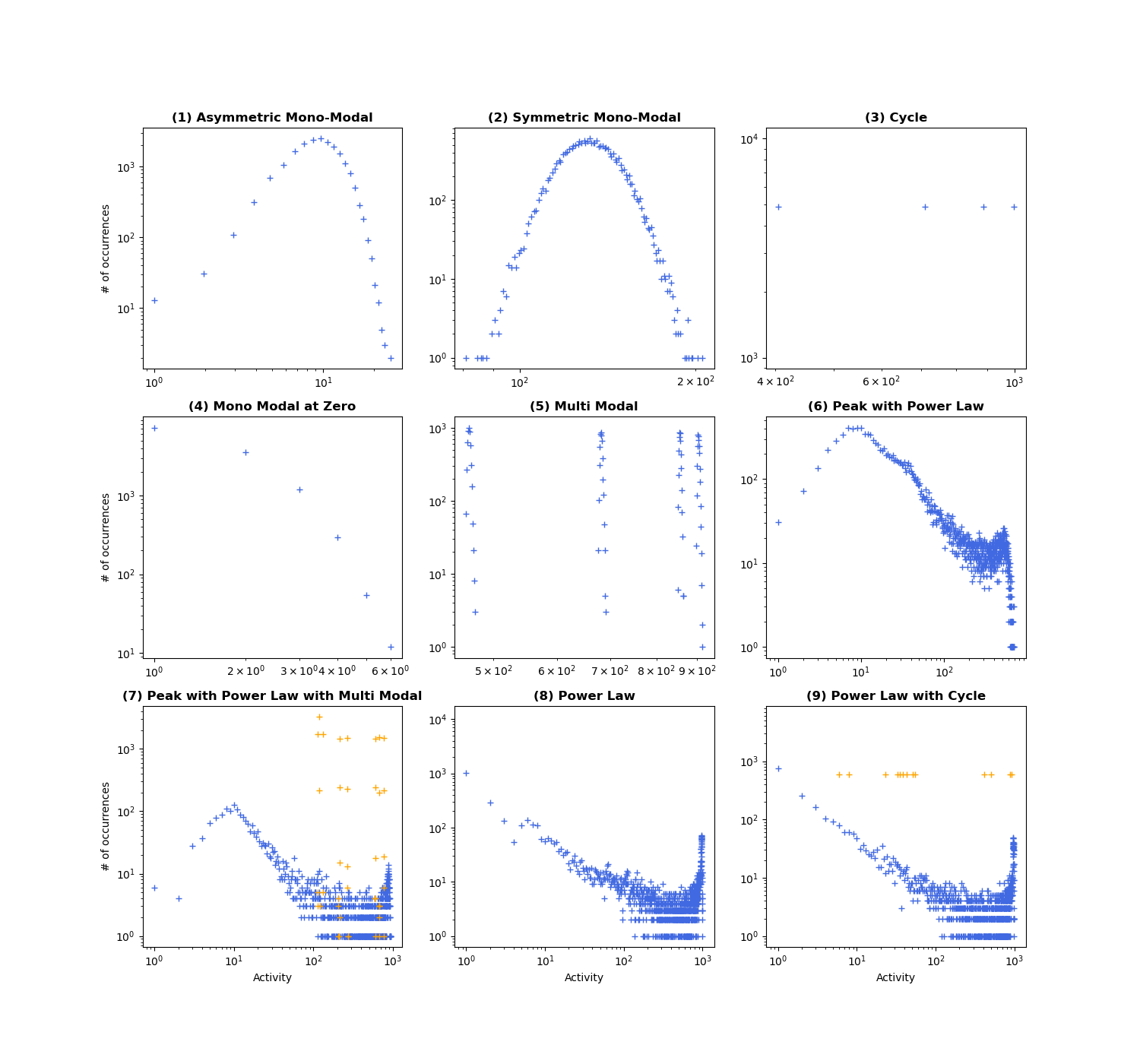}}
\caption{Qualitative analysis of SF network behaviors: (a) Average Activity plots over time and (b) Histograms of Total Activity
%for all the qualitative behaviors found in SF networks. 
Parameters for each panel: (1) \(k_0\) = 1, \(t_{ref}\) = 0, \(b\) = 2, \(p_{endo}\) = 0.01, and \(J\) = 1; (2) \(k_0\) = 1, \(t_{ref}\) = 0, \(b\) = 2, \(p_{endo}\) = 0.1, and \(J\) = 1; (3) \(k_0\) = 5, \(t_{ref}\) = 0, \(b\) = 3, \(p_{endo}\) = 0.01, and \(J\) = 2; (4) \(k_0\) = 1, \(t_{ref}\) = 0, \(b\) = 2, \(p_{endo}\) = 0.001, and \(J\) = 1; (5) \(k_0\) = 3, \(t_{ref}\) = 0, \(b\) = 3, \(p_{endo}\) = 0.1, and \(J\) = 1; (6) \(k_0\) = 3, \(t_{ref}\) = 10, \(b\) = 3, \(p_{endo}\) = 0.1, and \(J\) = 2; (7) \(k_0\) = 5, \(t_{ref}\) = 6, \(b\) = 3, \(p_{endo}\) = 0.1, and \(J\) = 2; (8) \(k_0\) = 5, \(t_{ref}\) = 10, \(b\) = 2, \(p_{endo}\) = 0.001, and \(J\) = 3; (9) \(k_0\) = 5, \(t_{ref}\) = 10, \(b\) = 2, \(p_{endo}\) = 0.01, and \(J\) = 3 \cite{cafiso_2024}.} \label{fig:SF_average_activity_total_activity}
\end{figure}

\noindent
We found a wide variety of dynamical behaviors that, surprisingly, are more common than expected between the two types of topologies.
The results of the qualitative parametric analysis are summarized in 
Figs. \ref{fig:ER_parametric_analysis} and \ref{fig:SF_parametric_analysis}
that present comprehensive parametric analysis results for the Erd$\ddot o$s-R\'eny and Scale-Free networks, respectively.
%for ER and SF networks respectively. 
The different qualitative behaviors are represented as point with different colours in a 3D plot with axes $k_0$, $\pi = \frac{J}{b}$ and $t_{ref}$.
%
%Figs. \ref{fig:ER_parametric_analysis} and \ref{fig:SF_parametric_analysis} present comprehensive parametric analysis results for the Erd$\ddot o$s-R\'eny and Scale-Free networks, respectively. 
Additionally, Figs. \ref{fig:ER_average_activity_total_activity} and \ref{fig:SF_average_activity_total_activity} illustrate the different kinds of activity patterns observed in each network connectivity type, providing insight into their distinctive dynamical properties.
Panels (a) in both figures show the temporal patterns of the average activity for the different dynamical behaviors that we have identified, ten typologies for the ER network and nine for for the SF network.
Panels (b) display the corresponding total activity histograms.
In particular, for the ER networks we have identified the following qualitative behaviors in the total activity distributions (see Fig. \ref{fig:ER_average_activity_total_activity}):
%\vspace{-.5cm}
\begin{itemize}
    \item[(1)] Asymmetric Mono-Modal distribution
    \item[(2)] Symmetric Mono-Modal distribution  
    \item[(3)] Mono-Modal and transition to multi-modal distribution     
    \item[(4)] Cycle distribution   
    \item[(5)] Mono-modal at zero distribution 
    \item[(6)] Multi-modal distribution
    \item[(7)] Peak with Power-Law distribution  
    \item[(8)] Peak with Power-Law and transition to multi-modal distribution 
    \item[(9)] Power-Law distribution
    \item[(10)] Power-Law with Multiple Peaks distribution    
\end{itemize}
%
\begin{comment}

\begin{enumerate}
    \item[(1)] Asymmetric Bell Curve distribution
    \item[(2)] Symmetric Bell Curve distribution  
    \item[(3)] Bell curve and transition to multi-modal distribution     
    \item[(4)] Cycle distribution   
    \item[(5)] Mono-modal at zero distribution 
    \item[(6)] Multi-modal distribution
    \item[(7)] Peak with Power-Law distribution  
    \item[(8)] Peak with Power-Law and transition to multi-modal distribution 
    \item[(9)] Power-Law distribution
    \item[(10)] Power-Law with Multiple Peaks distribution    
\end{enumerate}

\end{comment}
%
%%%%%%%%%%%%%%%%%%%%%%%%%%%%%%%%%%%
%%%%%%%%%%%%%%%%%%%%%%%%%%%%%%%%%%%

\noindent
For the SF networks we have identified the following qualitative behaviors in the total activity distributions (see Fig. \ref{fig:SF_average_activity_total_activity}):
% \vspace{-1.2cm}
\begin{itemize}
    \item[(1)] Asymmetric Mono-Modal distribution 
  % \vspace{-.3cm}      
    \item[(2)] Symmetric Mono-Modal distribution 
  % \vspace{-.3cm}  
    \item[(3)] Cycle distribution  
     % \vspace{-.3cm}  
    \item[(4)] Mono-modal at zero distribution 
     % \vspace{-.3cm}  
    \item[(5)] Multi-modal distribution 
     % \vspace{-.3cm}  
    \item[(6)] Peak with Power-Law distribution  
     % \vspace{-.3cm}  
    \item[(7)] Peak with Power-Law and transition to multi-modal distribution
     % \vspace{-.3cm}  
    \item[(8)] Power-Law distribution
     % \vspace{-.3cm}  
    \item[(9)] Power-Law and transition to cycle distribution
\end{itemize}
%

%\vspace{-1.cm}
%%%%%%%%%%%%%%%%%%%%%%%%%%%%%%%%%%%%%%%%%%%%%%%%%%%%%%%%%%%%%%
%%%%%%%% FIGURES PARAMETRIC ANALYSIS %%%%%%%%%%%%%%%%
%%%%%%%%%%%%%%%%%%%%%%%%%%%%%%%%%%%%%%%%%%%%%%%%%%%%%%%%%%%%%%
%
% Parametric analysis ER
%

%\vspace{-2.cm}
%\vspace{-1cm}
%
% Parametric analysis SF
%

%%%%%%%%%%%%%%%%%%%%%%%%%%%%%%%%%%%%%%%%%%%%%%%%%%%%%%%%%%%%%%
%%%%%%%%%%%%%%%%% FIGURES DYNAMICS ER/SF %%%%%%%%%%%%%%%%%%%%%
%%%%%%%%%%%%%%%%%%%%%%%%%%%%%%%%%%%%%%%%%%%%%%%%%%%%%%%%%%%%%%
%
% ER networks
%
%
% SF networks
%

\newpage

\subsection{Temporal Complexity Analysis}

\noindent
For each distinct dynamical pattern observed in both network types, we selected representative cases to investigate temporal complexity (TC) behaviors by applying the EDDiS algorithm to the neural coincidence event sequences. Comprehensive details of this analytical methodology are provided in Appendix \ref{appendix:eddis}.

\begin{figure}[H]
\centering
\subfloat[DFA]{\includegraphics[scale=0.25]{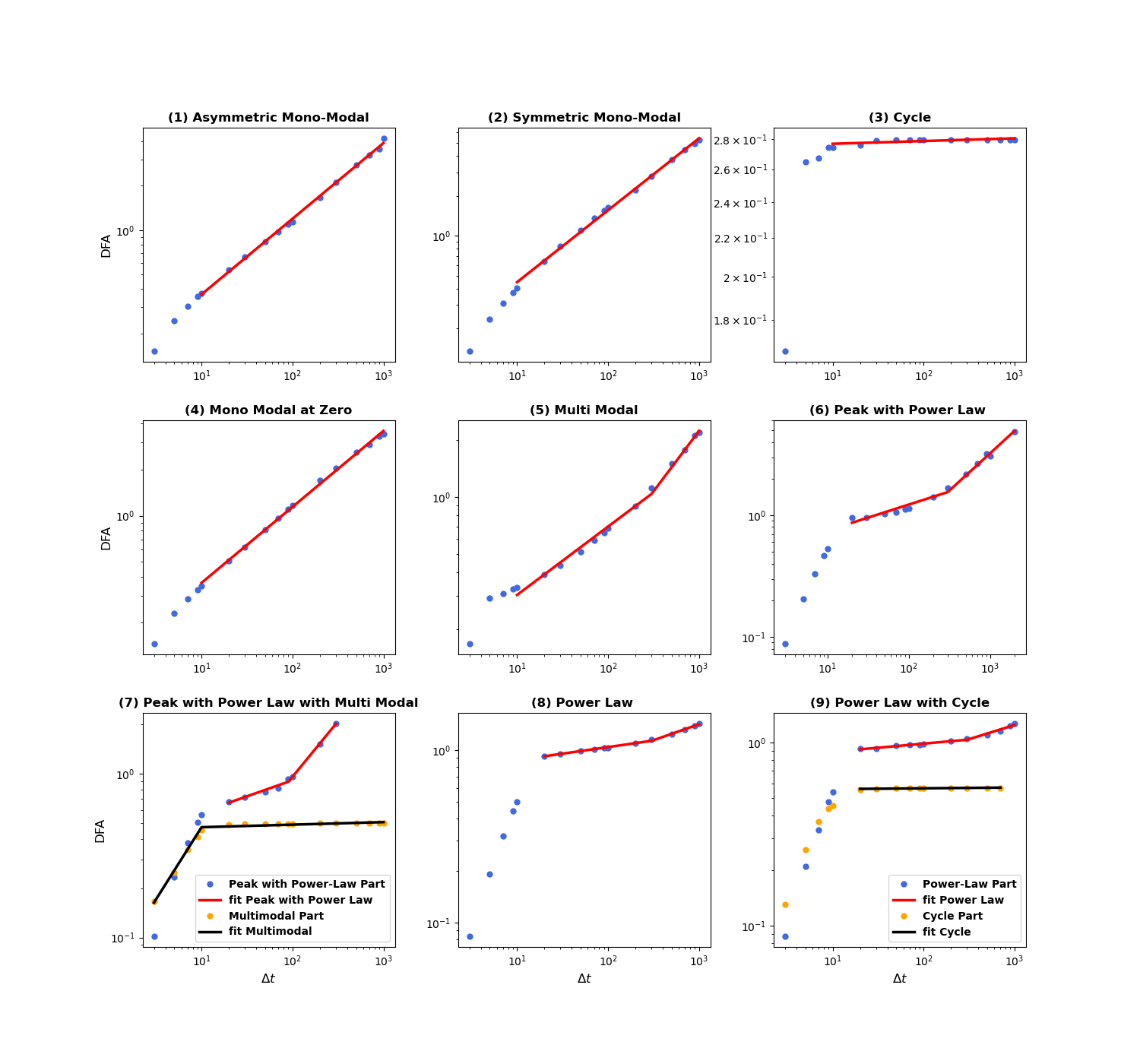}}
\hspace{5.cm}
\subfloat[DE]{\includegraphics[scale=0.25]{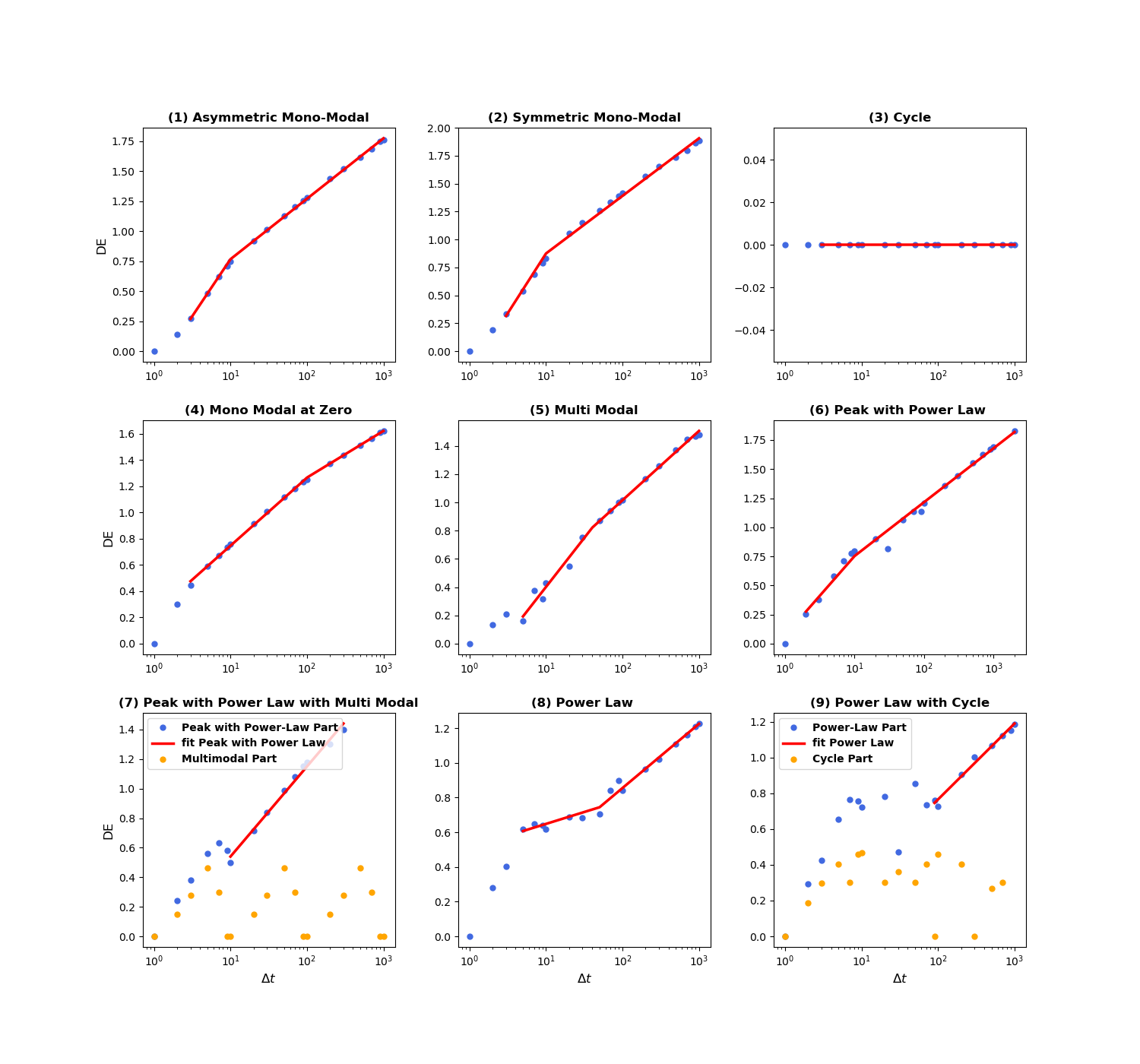}}
\caption{
(a) DFA and (b) DE plots with fits for all the qualitative behaviors found in SF networks. Parameters for each panel: (1) \(k_0\) = 1, \(t_{ref}\) = 0, \(b\) = 2, \(p_{endo}\) = 0.01, and \(J\) = 1; (2) \(k_0\) = 1, \(t_{ref}\) = 0, \(b\) = 2, \(p_{endo}\) = 0.1, and \(J\) = 1; (3) \(k_0\) = 5, \(t_{ref}\) = 0, \(b\) = 3, \(p_{endo}\) = 0.01, and \(J\) = 2; (4) \(k_0\) = 1, \(t_{ref}\) = 0, \(b\) = 2, \(p_{endo}\) = 0.001, and \(J\) = 1; (5) \(k_0\) = 3, \(t_{ref}\) = 0, \(b\) = 3, \(p_{endo}\) = 0.1, and \(J\) = 1; (6) \(k_0\) = 3, \(t_{ref}\) = 10, \(b\) = 3, \(p_{endo}\) = 0.1, and \(J\) = 2; (7) \(k_0\) = 5, \(t_{ref}\) = 6, \(b\) = 3, \(p_{endo}\) = 0.1, and \(J\) = 2; (8) \(k_0\) = 5, \(t_{ref}\) = 10, \(b\) = 2, \(p_{endo}\) = 0.001, and \(J\) = 3; (9) \(k_0\) = 5, \(t_{ref}\) = 10, \(b\) = 2, \(p_{endo}\) = 0.01, and \(J\) = 3.} 
\label{fig:SF_DFA_DE}
\end{figure}

\noindent
Figs. \ref{fig:ER_DFA_DE} present the Detrended Fluctuation Analysis (DFA) and Diffusion Entropy (DE) results for ER networks, while Figs. \ref{fig:SF_DFA_DE} show corresponding analyses for SF networks. Tables \ref{tab:results_DFA_DE_ER} and \ref{tab:results_DFA_DE_SF} summarize the best-fit values of scaling exponents $H$ and $\delta$ for all dynamical patterns in ER and SF networks, respectively. 

\noindent
The DFA results reveal distinct scaling behaviors across activity patterns.  Asymmetric Mono-Modal, Symmetric Mono-Modal and Mono-Modal at Zero patterns (including the bell-shaped component in Mono-Modal with Multi-Modal cases) exhibit $H \sim 0.5$, indicating normal diffusion. 
According to EDDiS algorithm, this means that the total activity pattern can be described as a random signal with no or short correlations.
Power-Law cases, including those with activity transitions and multiple peaks, show $H \sim 0$ in short-time regimes. Conversely, these same cases show a transition to values $H \sim 0.2$ in long-time regimes, demonstrating subdiffusive behavior and, thus, anti-correlated patterns. 
This means that the system  converges to patterns that remain almost stable in time.
Cycle cases, also when in transition with a Power-Law behavior, and the Multi-Modal component of Bell-curve with Multi-Modal patterns display $H = 0$, suggesting no diffusion. 
The limit case $H \sim 0$ over all times means that the system is totally trapped in deterministic patterns. The case of cycles is prototypical of this behavior.
Peak with Power-Law patterns show network-dependent behavior: in SF networks, they exhibit subdiffusive scaling $H \sim 0.2$  in short-time regimes and a crossover to superdiffusive $H \sim 0.6$ in long-time regimes. This implies a transition from anti-correlated patterns to slightly correlated one, i.e., from a quasi-trapped phase to a long time regime where the system has a weak tendency to explore all the possible configurations.
%, suggesting a shift from subdiffusive to more complex behavior. 
In ER networks, these patterns display $H \sim 0.1$ in short-time regimes and $H \sim 0.3$ in long time regimes (subdiffusive behavior), while transitions with Multi-Modal patterns show $H \sim 0.2$ to $H \sim 0.5$  (subdiffusive to normal behavior). The Multi-Modal case, in particular,  presents a wide variety of behaviors.
In SF networks, pure Multi-Modal patterns show $H \sim 0.4$ in short-time regimes shifting to $H \sim 0.6$ in long-time regimes, while transitions with Peak with Power-Law patterns display $H = 0.88$ in short-time regimes decreasing to $H \sim 0$ in long-time regimes. In ER networks pure Multi-Modal pattern exhibit $H \sim 0.2$ in short-time regimes increasing to $H > 1$ in long-time regimes, indicating a transition from trapped patterns to positively correlated temporal patterns of total network activity.
%from subdiffusive to superdiffusive behavior.

\noindent
DE analysis provides complementary insights. Mono-Modal patterns (both symmetric and asymmetric), including transitions with Multi-Modal patterns, show $\delta \sim 0.5$ in long -time regimes (normal diffusion), while exhibiting $\delta \sim 1$ in SF networks and $\delta \sim 0.8$ in ER networks during short-time regimes. Mono-Modal at Zero patterns in SF networks display $\delta \sim 0.5$ in  short-time regimes decreasing to $\delta \sim 0.3$ in long-time regimes, while in ER networks they maintain $\delta \sim 0.4$ in long-time regimes. Power-Law behaviors consistently show $\delta = 0.4$ across both network types, indicating subdiffusive behavior. Cycle cases and certain Multi-Modal transitions exhibit $\delta = 0$, confirming no diffusion. Peak with Power-Law patterns generally show $\delta \sim 0.6$  in short-time regimes transitioning to $\delta \sim 0.4$ in long-time regimes, whereas Multi-Modal cases typically display either $\delta \sim 0$ or $\delta \sim 0.4$ , indicating either no diffusion or subdiffusive behavior.

\begin{figure}[H]
\centering
\subfloat[DFA]{\includegraphics[scale=0.24]{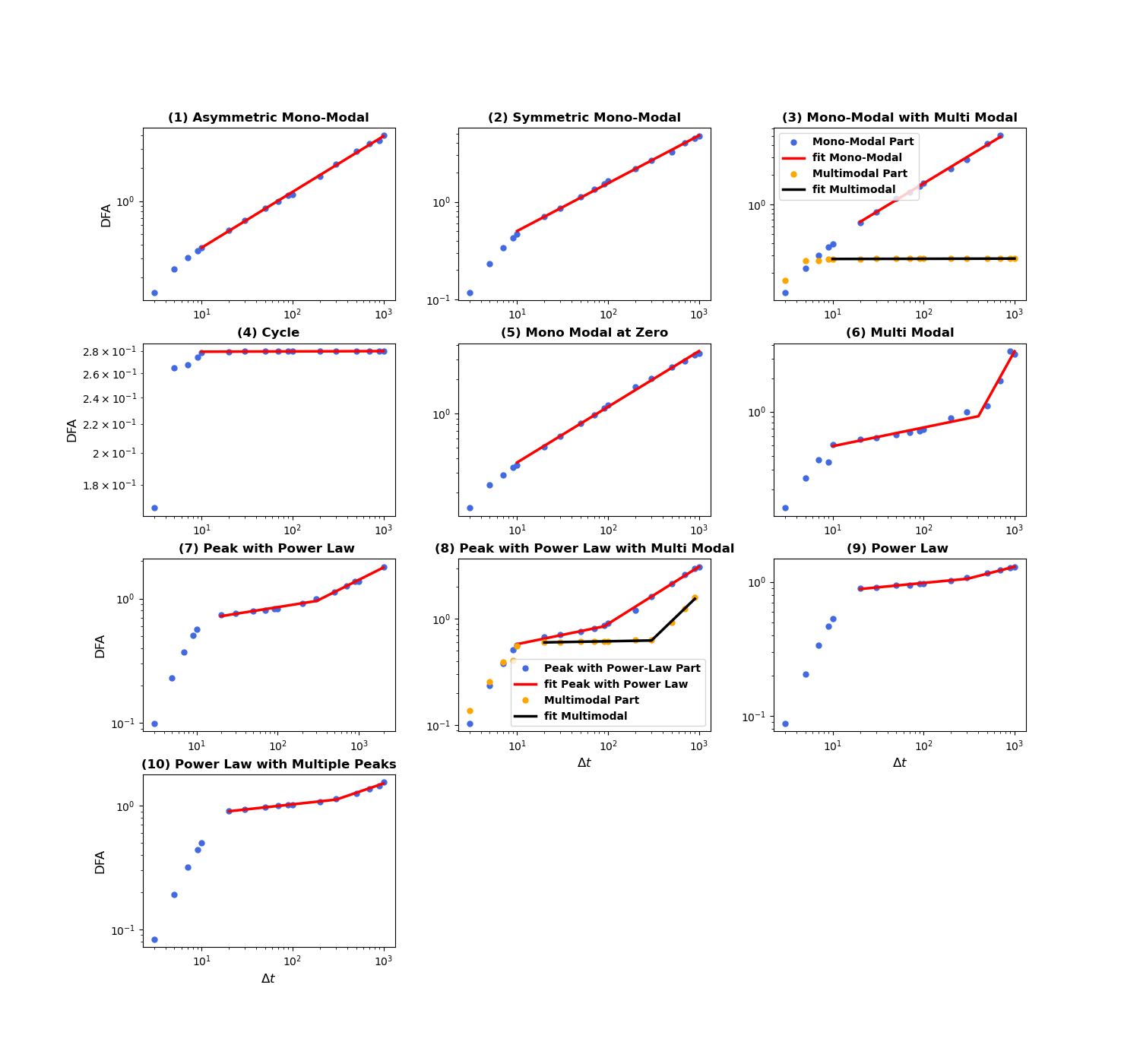}}
\hspace{5.cm}
\subfloat[DE]{\includegraphics[scale=0.25]{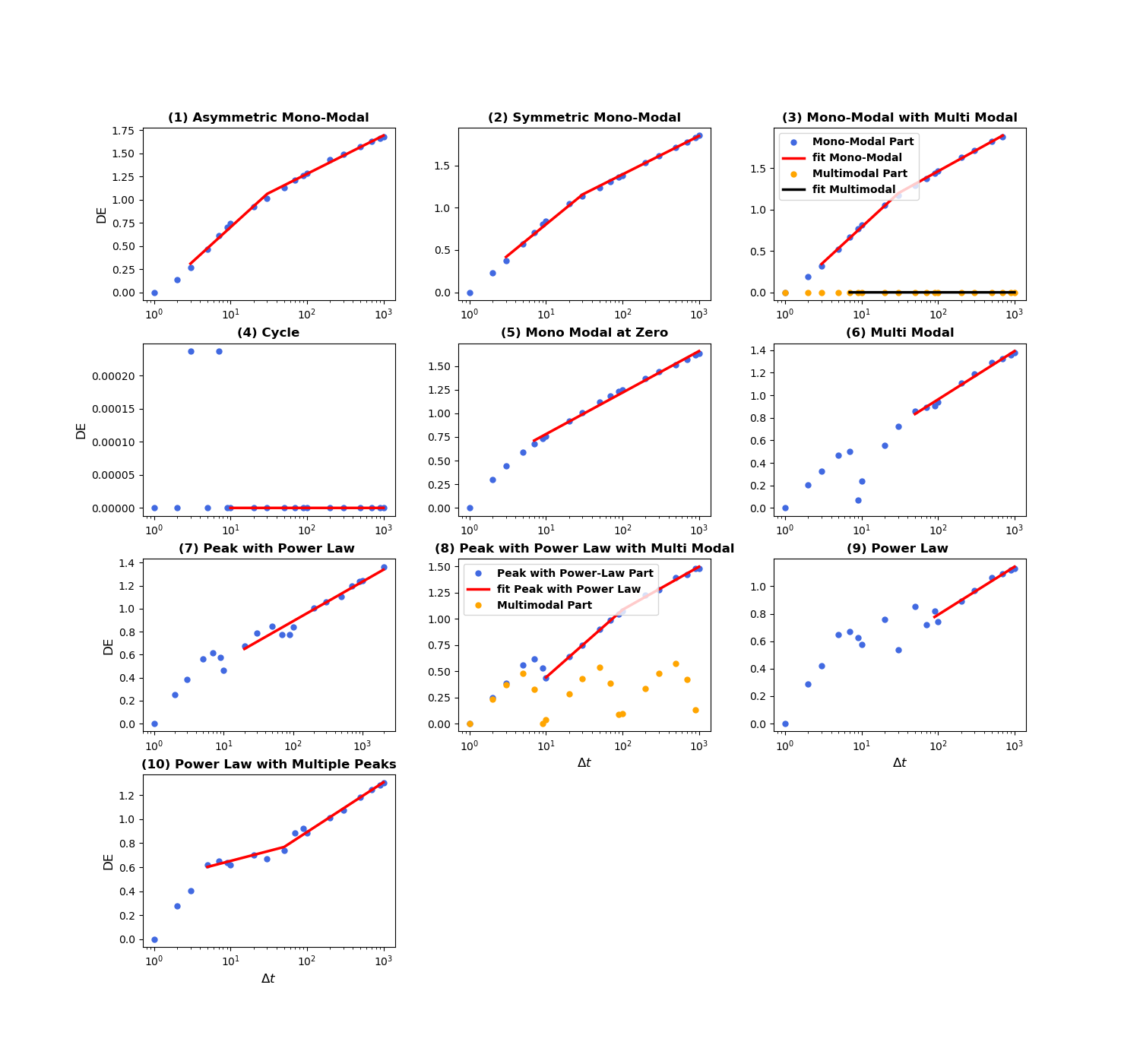}}
\caption{(a) DFA and (b) DE plots with fits for all the qualitative behaviors found in ER networks. Parameters for each panel: (1) \(k_0\) = 1, \(t_{ref}\) = 4, \(b\) = 2, \(p_{endo}\) = 0.01, and \(J\) = 1; (2) \(k_0\) = 1, \(t_{ref}\) = 4, \(b\) = 2, \(p_{endo}\) = 0.1, and \(J\) = 1;  (3) \(k_0\) = 2, \(t_{ref}\) = 0, \(b\) = 2, \(p_{endo}\) = 0.01, and \(J\) = 1; (4) \(k_0\) = 4, \(t_{ref}\) = 0, \(b\) = 2, \(p_{endo}\) = 0.01, and \(J\) = 1; (5) \(k_0\) = 1, \(t_{ref}\) = 4, \(b\) = 2, \(p_{endo}\) = 0.001, and \(J\) = 1; (6) \(k_0\) = 3, \(t_{ref}\) = 6, \(b\) = 3, \(p_{endo}\) = 0.1, and \(J\) = 2; (7) \(k_0\) = 5, \(t_{ref}\) = 6, \(b\) = 3, \(p_{endo}\) = 0.1, and \(J\) = 1; (8) \(k_0\) = 5, \(t_{ref}\) = 6, \(b\) = 3, \(p_{endo}\) = 0.1, and \(J\) = 2; (9) \(k_0\) = 5, \(t_{ref}\) = 10, \(b\) = 2, \(p_{endo}\) = 0.01, and \(J\) = 3; (10) \(k_0\) = 5, \(t_{ref}\) = 10, \(b\) = 2, \(p_{endo}\) = 0.001, and \(J\) = 3.}
\label{fig:ER_DFA_DE}
\end{figure}
%
% SF networks
%

%
% ER-Networks DFA/DE
%
\begin{table}[h]
        \caption{Result of fits for DFA and DE for ER Networks. PPL: Peak with Power-Law. MuM: Multi-Modal. MoM: Mono-Modal.
%        Result of fits for DFA and DE for ER Networks. For the ``Peak with Power-Law with Multi-Modal'' row: Peak with Power-Law part is indicated with (PPL), and Multi-Modal part is indicated with (MM). For the ``Bell Curve with Multi-Modal'' row: Bell Curve part is indicated with (BC), and Multi-Modal part is indicated with (MM). 
        Simulation parameters for each case: (1) \(k_0\) = 1, \(t_{ref}\) = 4, \(b\) = 2, \(p_{endo}\) = 0.01, and \(J\) = 1; (2) \(k_0\) = 1, \(t_{ref}\) = 4, \(b\) = 2, \(p_{endo}\) = 0.1, and \(J\) = 1;  (3) \(k_0\) = 2, \(t_{ref}\) = 0, \(b\) = 2, \(p_{endo}\) = 0.01, and \(J\) = 1; (4) \(k_0\) = 4, \(t_{ref}\) = 0, \(b\) = 2, \(p_{endo}\) = 0.01, and \(J\) = 1; (5) \(k_0\) = 1, \(t_{ref}\) = 4, \(b\) = 2, \(p_{endo}\) = 0.001, and \(J\) = 1; (6) \(k_0\) = 3, \(t_{ref}\) = 6, \(b\) = 3, \(p_{endo}\) = 0.1, and \(J\) = 2; (7) \(k_0\) = 5, \(t_{ref}\) = 6, \(b\) = 3, \(p_{endo}\) = 0.1, and \(J\) = 1; (8) \(k_0\) = 5, \(t_{ref}\) = 6, \(b\) = 3, \(p_{endo}\) = 0.1, and \(J\) = 2; (9) \(k_0\) = 5, \(t_{ref}\) = 10, \(b\) = 2, \(p_{endo}\) = 0.01, and \(J\) = 3; (10) \(k_0\) = 5, \(t_{ref}\) = 10, \(b\) = 2, \(p_{endo}\) = 0.001, and \(J\) = 3}
        \begin{center}
        \begin{tabular}{c|c|c|c|c|}
            \cline{2-5}
             & \multicolumn{2}{|c|}{DFA ($H$)} & \multicolumn{2}{|c|}{DE ($\delta$)}\\
            \cline{2-5}
             & Short-Time & Long-Time & Short-Time & Long-Time \\
            \hline
            \multicolumn{1}{|c|}{(1) Asymmetric Mono-Modal} & 
            / & 
            0.51 &
            0.75 & 
            0.41 \\
            \hline  
            \multicolumn{1}{|c|}{(2) Symmetric Mono-Modal} & 
            / & 
            0.49 &
            0.74 & 
            0.45 \\
            \hline
            \multicolumn{1}{|c|}{(3) Mono-Modal Curve with Multi-Modal} &
            %Bell Curve with Multi-Modal} &  
            / (MoM) &  
            0.56 (MoM) &
            0.85 (MoM) & 
            0.51 (MoM) \\
            \multicolumn{1}{|c|}{(Non-Stationary)} & 
            / (MuM) &
            0 (MuM) &
            / (MuM) &
            0 (MuM) \\
            \hline
            \multicolumn{1}{|c|}{(4) Cycle} & 
            / & 
            0 & 
            / & 
            0 \\
            \hline
            \multicolumn{1}{|c|}{(5) Mono-Modal at Zero} & 
            / & 
            0.49 &
            / & 
            0.44 \\
            \hline
            \multicolumn{1}{|c|}{(6) Multi-Modal} & 
            0.17 &
            1.47 & 
            / & 
            0.43 \\
            \hline
            \multicolumn{1}{|c|}{(7) Peak with Power-Law} & 
            0.10 & 
            0.33 & 
            / & 
            0.34 \\
            \hline
            \multicolumn{1}{|c|}{(8) Peak with Power-Law with Multi-Modal } &  
            0.17 (PPL) &
            0.54 (PPL) &
            0.64 (PPL) &
            0.41 (PPL) \\
            \multicolumn{1}{|c|}{(Non-stationary)} & 
            0.01 (MuM) &
            0.82 (MuM) &
            / (MuM) &
            0.07 (MuM)\\
            \hline             
            \multicolumn{1}{|c|}{(9) Power-Law} & 
            0.06 & 
            0.17 & 
            / & 
            0.35 \\
            \hline
             \multicolumn{1}{|c|}{(10) Power-Law with Multiple Peaks} &  
            0.08 &  
            0.25 &
            0.16 &
            0.41\\
            \hline          
        \end{tabular}
        \end{center}
    \label{tab:results_DFA_DE_ER}
\end{table}

%
% SF-Networks DFA/DE
%
\begin{table}[h]
        \caption{Result of fits for DFA and DE for SF Networks. PL: Power-Law. C: Cycle. PPL: Peak with Power-Law. MuM: Multi-Modal.
%        Result of fits for DFA and DE for SF Networks. For the ``Power-Law with Cycle'' row: Power-Law part is indicated with (PL), and Cycle part is indicated with (C). For the ``Peak with Power-Law with Multi-Modal'' row: Peak with Power-Law part is indicated with (PPL), and Multi-Modal part is indicated with (MM). 
        Simulation parameters for each case: (1) \(k_0\) = 1, \(t_{ref}\) = 0, \(b\) = 2, \(p_{endo}\) = 0.01, and \(J\) = 1; (2) \(k_0\) = 1, \(t_{ref}\) = 0, \(b\) = 2, \(p_{endo}\) = 0.1, and \(J\) = 1; (3) \(k_0\) = 5, \(t_{ref}\) = 0, \(b\) = 3, \(p_{endo}\) = 0.01, and \(J\) = 2; (4) \(k_0\) = 1, \(t_{ref}\) = 0, \(b\) = 2, \(p_{endo}\) = 0.001, and \(J\) = 1; (5) \(k_0\) = 3, \(t_{ref}\) = 0, \(b\) = 3, \(p_{endo}\) = 0.1, and \(J\) = 1; (6) \(k_0\) = 3, \(t_{ref}\) = 10, \(b\) = 3, \(p_{endo}\) = 0.1, and \(J\) = 2; (7) \(k_0\) = 5, \(t_{ref}\) = 6, \(b\) = 3, \(p_{endo}\) = 0.1, and \(J\) = 2; (8) \(k_0\) = 5, \(t_{ref}\) = 10, \(b\) = 2, \(p_{endo}\) = 0.001, and \(J\) = 3; (9) \(k_0\) = 5, \(t_{ref}\) = 10, \(b\) = 2, \(p_{endo}\) = 0.01, and \(J\) = 3}
        \begin{center}
        \begin{tabular}{c|c|c|c|c|}
            \cline{2-5}
             & \multicolumn{2}{|c|}{DFA ($H$)} & \multicolumn{2}{|c|}{DE ($\delta$)}\\
            \cline{2-5}
             & Short-Time & Long-Time & Short-Time & Long-Time \\
            \cline{2-5}
            \hline
            \multicolumn{1}{|c|}{(1) Asymmetric Mono-Modal} & 
            / & 
            0.51 & 
            0.95 & 
            0.50 \\
            \hline
            \multicolumn{1}{|c|}{(2) Symmetric Mono-Modal} & 
            / & 
            0.54 &
            1.07 & 
            0.51 \\
            \hline
            \multicolumn{1}{|c|}{(3) Cycle} & 
            / & 
            0 & 
            / & 
            0 \\
            \hline     
            \multicolumn{1}{|c|}{(4) Mono-Modal at Zero} & 
            / & 
            0.50 & 
            0.52 & 
            0.35 \\
            \hline
            \multicolumn{1}{|c|}{(5) Multi-Modal} & 
            0.36 & 
            0.64 & 
            0.69 & 
            0.49 \\
            \hline
            \multicolumn{1}{|c|}{(6) Peak with Power-Law} & 
            0.21 & 
            0.61 & 
            0.68 & 
            0.46 \\
            \hline
            \multicolumn{1}{|c|}{(7) Peak with Power-Law with Multi-Modal } &  
            0.19 (PPL) & 
            0.68 (PPL) &
            / (PPL) &
            0.61 (PPL) \\
            \multicolumn{1}{|c|}{} & 
            0.88 (MuM) &  
            0.01 (MuM) &
            / (MuM) &
            0 (MuM)\\
            \hline            
            \multicolumn{1}{|c|}{(8) Power-Law} & 
            0.07 &
            0.18 &
            0.17 & 
            0.37 \\
            \hline
            \multicolumn{1}{|c|}{(9) Power-Law with Cycle} &  
            0.04 (PL) &  
            0.15 (PL) &
            / (PL) &
            0.42 (PL)\\
            \multicolumn{1}{|c|}{} & 
            / (C) &  
            0 (C) &
            / (C) &
            / (C)\\
            \hline            
        \end{tabular}
        \end{center}
    \label{tab:results_DFA_DE_SF}
\end{table}

%%%%%%%%%%%%%%%%%%%%%%%%%%%%%%%%%%%%%%%%%%%%%%%%%%%%%%%%%%%%%%%
\section{Discussion}
\label{sec:discussion}

Our computational simulations revealed diverse behavioral patterns in the Hopfield-type network across both network architectures, namely  Erd$\ddot o$s-R\'eny (ER) and scale-free (SF).
As illustrated in Figs. \ref{fig:ER_average_activity_total_activity} and \ref{fig:SF_average_activity_total_activity}, most qualitative dynamics appear in both network architectures, even with different parameter configurations. In particular, the 'Mono Modal at 0' pattern (panels (4) in Fig. \ref{fig:SF_average_activity_total_activity} and (5) in Fig. \ref{fig:ER_average_activity_total_activity}) appears consistently within the same parameter ranges across both network types. Similarly, 'Asymmetric Mono-Modal' patterns (panels (1) in Fig. \ref{fig:SF_average_activity_total_activity} and \ref{fig:ER_average_activity_total_activity}) emerge under comparable parameter conditions in both topological configurations. Notable differences include the exclusive presence of the "Mono-Modal with Multi-Modal" pattern (panels (3) in \ref{fig:ER_average_activity_total_activity})  in ER networks, observed only under two specific parameter configurations, and the occurrence of several 'Multi-Modal' patterns (panels (5) in Fig. \ref{fig:SF_average_activity_total_activity})  in SF networks. 'Symmetric Mono-Modal' patterns (panels (2) in Fig. \ref{fig:SF_average_activity_total_activity} and \ref{fig:ER_average_activity_total_activity})  manifest in both neural network types exclusively when the endogenous firing probability is elevated ($p_{endo} = 0.1$). Increasing the minimum out-degree value and coupling parameter appears to transform a pure 'Symmetric Bell-Curve' into a 'Peak with Power-Law' pattern (panels (6) in Fig. \ref{fig:SF_average_activity_total_activity} and (7) in Fig. \ref{fig:ER_average_activity_total_activity}), which dominates within those parameter ranges. The 'Multi-Modal' patterns (panels (5) in Fig. \ref{fig:SF_average_activity_total_activity} and (6) in Fig. \ref{fig:ER_average_activity_total_activity})  occur across numerous parameter configurations in ER networks but appear significantly less often in SF network configurations. In SF networks, these patterns primarily emerge under conditions of fewer network links combined with strong inter-neuronal coupling. Conversely, 'Cycle' patterns(panels (3) in Fig. \ref{fig:SF_average_activity_total_activity} and (4) in Fig. \ref{fig:ER_average_activity_total_activity})  predominate in SF networks, likely attributable to the hub structures inherent to this topological configuration. Pure 'Power-Law' patterns  (panels (8) in Fig. \ref{fig:SF_average_activity_total_activity} and (9) in Fig. \ref{fig:ER_average_activity_total_activity}) emerge in both topological configurations within similar parameter ranges but require different endogenous firing probabilities. SF networks apparently require much lower endogenous firing probabilities compared to ER networks to exhibit pure 'Power-Law' behaviors. 
This is further evidenced by the presence of 'Power-Law with Multiple Peaks' patterns (panels (10) in Fig. \ref{fig:ER_average_activity_total_activity})—a distinctive feature of ER networks—occurring within the same parameter ranges as pure 'Power-Law' behaviors in SF networks. 'Peak with Power-Law with Multi-Modal' patterns (panels (7) in Fig. \ref{fig:SF_average_activity_total_activity} and (8) in Fig. \ref{fig:ER_average_activity_total_activity}) occur with substantially higher frequency in ER networks relative to SF networks. Finally, the 'Power-Law with Cycle' pattern (panels (9) in Fig. \ref{fig:SF_average_activity_total_activity}), unique to SF networks, appears exclusively under conditions of low endogenous firing probability, particularly in denser networks with extended refractory periods and strong coupling parameters. 

\noindent
%
% EDDiS Results
%
As anticipated above, the estimations of temporal complexity levels of the networks were done by applying the EDDiS analysis to one specific sample for each activity pattern found in the parametric analysis. For dynamic cases displaying clear transitions in activity patterns, we implemented the EDDiS algorithm with separate analyses of distinct dynamical behaviors. Our investigation revealed diverse diffusive characteristics across the identified dynamical patterns.
The DFA results indicate that 'Asymmetric Mono-Modal', 'Symmetric Mono-Modal', and 'Mono-Modal at Zero' patterns exhibit normal diffusive behavior with Hurst exponent values approximately equal to $0.5$ ($H\sim0.5$). Similarly, the 'Bell-Curve' component of the case 'Mono-Modal with Multi-Modal' maintains $H \sim 0.5$, confirming normal diffusive behavior. As expected, given its redundant behavior, 'Cycle' patterns consistently demonstrate $H = 0$ values, even when transitioning with 'Power-Law' behavior. In ER networks, all 'Power-Law' patterns—pure or hybrid—exhibited Hurst exponent values less than or equal to $0.5$ ($H\leq0.5$), indicating strong anti-correlation in the total activity temporal patterns. In SF networks the 'Power-Law', pure or with a transition to 'Cycle', also showed subdiffusive characteristics ($H < 0.5$), consistent with ER networks. However, 'Peak with Power-Law' patterns—including those transitioning with 'Multi-Modal' behavior—displayed $H < 0.5$ at short time-lags and $0.6<H<0.7$ at long time-lags, suggesting a transition between 
an anti-correlated and a correlated temporal pattern.
%
%subdiffusive and potentially complex behaviors (with $\mu \sim %1.3$ or $\mu \sim 2.7$). 
The 'Multi-Modal' pattern, being the most dynamically unstable, exhibited $H < 0.5$ at short time-lags in both network types. At long time-lags, however, it showed $H > 1$ in ER networks, indicating strong correlation associated with superdiffusive behavior, while SF networks displayed $H \sim 0.6$, suggesting possible complex behavior (with $\mu \sim 1.2$ or $\mu \sim 2.4$). The DE analysis corroborates the diffusive behaviors identified through DFA across all examined cases. Despite its simplicity, this model generates a rich spectrum of complex scenarios predominantly characterized by subdiffusive or normal diffusive behaviors, with selected instances potentially exhibiting superdiffusive regimes and, thus, strong correlated patterns of network total activity.

%%%%%%%%%%%%%%%%%%%%%%%%%%%%%%%%%%%%%%%%%%%%%%%%%%%%%%%%%%%%%%%
\section{Concluding remarks}
\label{sec:conclusion}

In this study, we have examined a Hopfield-type neural network model, specifically the framework presented in \cite{grinstein2005model}, which serves as a fundamental example of biologically-inspired neural architecture. This model incorporates several biologically-relevant characteristics, including adjustable refractory periods and maximum firing durations that can be calibrated and conceptualized as hyperparameters within artificial intelligence applications. The Hebbian learning mechanism inherent to Hopfield-type networks represents another significant biologically-inspired attribute, functioning as an unsupervised learning paradigm. Network architecture has been identified in recent literature as a crucial factor influencing both the collective network dynamics and learning efficacy. This architectural influence is particularly significant within the artificial intelligence domain. Various researchers have investigated how topological organization affects learning capabilities \cite{kaviani_eswa2021}, with some studies demonstrating enhanced learning performance associated with particular network configurations, such as scale-free and/or small-world structures \cite{Lu_Chaos2023}.

\noindent
This study constitutes an initial comprehensive exploration of the connections between structural patterns and temporal complexity in a basic spiking neural network lacking adaptive learning mechanisms. Notably, diverse architectural configurations can produce comparable dynamic patterns and similar complexity characteristics. Most of the dynamical behaviors found seems to be related to subdiffusive or normal diffusive behaviors and only certain patterns can be related to a superdiffusive behavior. Thus, most parameter settings lead to strongly anti-correlated activity patterns, while a relatively small region in the parametric space leads to correlate patterns of network activity and, thus, to a more extended exploration of the configuration space by the network dynamics.
The conditions compatible with a higher temporal complexity are not only associated with the emergence of a power-law decay in the total activity distribution, but also with other emerging statistics such as the Multi-Modal case.
Interestingly, most of the superdiffusive behaviors were found in networks with scale-free topology.

\noindent
Concerning the connection between network architecture and temporal complexity, additional research is necessary to clarify why distinctly different structural organizations can generate comparable complexity patterns. 
Future studies need to be carried out to deepen our comprehension of the interrelationships among the network's dynamic properties (such as temporal complexity), connectivity configuration, and learning characteristics including information storage capacity.

%\noindent
%Additionally, we intend to conduct subsequent investigations examining the interplay between connectivity structure and learning algorithms, to analyze how performance metrics, together with complexity measure assessments, evolve as the quantity of stored patterns increases (for example, in Hopfield model implementations).

%%%%%%%%%%%%%%%%%%%%%%%%%%%%%%%%%%%%%%%%%%%%%%%%%%%%%%%%%%
%%%%%%%%%%%%%%%%%%%%%% APPENDIX %%%%%%%%%%%%%%%%%%%%%%%%%% 
\appendix

%%%%%%%%%%%%%%%%%%%%%%%%%%%%%%%%%%%%%%%%%%%%%%%%%%%%%%%%%%
%%%%%%%%%%%%%%%%%%%%%%%%%%%%%%%%%%%%%%%%%%%%%%%%%%%%%%%%%%
\section{Event-driven diffusion scaling Analysis}
\label{appendix:eddis}

The diffusion scaling analysis is a robust technique for scaling detection, and when applied to a series of transition events, it can provide valuable insights into the underlying dynamics that actually produce these events.
The TC analysis utilizes the Event-Driven Diffusion Scaling (EDDiS) algorithm 
\cite{allegrini_pre2009,paradisi_csf15_pandora,paradisi_springer2017},
which involves calculating three distinct random walks by applying different walking rules to a sequence of observed transition events. This process includes computing the second moment scaling and the similarity of the diffusion probability distribution function (PDF). 
The underlying concept is rooted in the Continuous Time Random Walk (CTRW) model \cite{montroll1964random,weiss1983random}, where particle movement occurs only at event occurrence times.
Specifically, we focus on the Asymmetric Jump (AJ) walking rule \cite{grigolini2001asymmetric}, which involves making a single-step jump forward upon each event occurrence, effectively mirroring the counting process generated by the event sequence:
\beq
X(t) = \#\{n: t_n < t\}.
\label{ctrw_aj}
\eeq
The method used to extract global events from the simulated data and the scaling analyses are described in the following.

%%%%%%%%%%%%%%%%%%%%%%%%%%%%%%%%%%%%%%%%%%%%%%%%%%%%%%
\subsection{Neural coincidence events}
\label{app:neural_coinc}

The TC features here investigated are applied to {\it coincidence events}, which are defined as instances where at least a minimum number $N_c$ of neurons fire simultaneously. 
Given the comprehensive set of individual neuron firing times, the time of a coincidence event is determined as the moment when more than $N_c$ nodes fire together, i.e., within a tolerance interval of duration $\Delta t_c$.
In this work, we set $\Delta t_c$ equal to a single sampling period, i.e., $\Delta t_c = 1$, which effectively means we are looking for simultaneous events.
The overall activity distribution of the network corresponds to the size distribution of coincidences with a minimum threshold of $N_c=1$: $P(n_c| N_c=1)$. The actual threshold $N_c$ used here is determined by calculating the $35$th percentile of $P(n_c| N_c=1)$. Each $n$-th synchronous event is characterized by its occurrence time $t_c(n)$ and its size $S_c(n)$.

%%%%%%%%%%%%%%%%%%%%%%%%%%%%%%%%%%%%%%%%%%%%%%%%%%%%
\subsection{Detrended Fluctuation Analysis (DFA)}

\noindent
DFA is a well-known algorithm (see, e.g., \cite{peng_pre94}) that is widely used in the literature for the evaluation of  the second-moment scaling $H$ defined by:
\begin{eqnarray}
    &F^{^2}(\Delta t) = \langle \left( \Delta X(\Delta t) - \Delta X_{\rm trend}(\Delta t) \right )^2 \rangle \sim t^{2H}
    \label{h_scaling}\\
    \ \nonumber \\
    & F(\Delta t) = a \cdot \Delta t^H  \Rightarrow \nonumber \\
    & \Rightarrow  \log(F(\Delta t)) = \log(a)+ H \cdot \log(\Delta t) 
    \label{dfa_function}
\end{eqnarray}
being $\Delta X (t,\Delta t) = X(t+\Delta t) - X(t)$.
We denote the scaling exponent as $H$ because it is fundamentally equivalent to the classical Hurst similarity exponent, as introduced by Hurst \cite{hurst_1951}. The term $X_{\rm trend}(\Delta t)$ represents a suitable local trend within the time series. The DFA is calculated across various time lags $\Delta t$, and the statistical average is obtained by dividing the time series into multiple time windows of duration $\Delta t$.
In the EDDiS method, the DFA is applied to different event-driven diffusion processes, as detailed in works such as \cite{paradisi_springer2017,paradisi_npg12}. To verify the accuracy of Eq. \eqref{dfa_function} for the data and to estimate the exponent $H$, it is sufficient to perform a linear best fit in the logarithmic scale.
To perform the DFA, we utilized the MFDFA function from the MFDFA package in Python \cite{gorjao2022MFDFA}.

%%%%%%%%%%%%%%%%%%%%%%%%%%%%%%%%%%%%%%%%%%%%%%%%%%%%%%%%%%%%%
\subsection{Diffusion Entropy}

%Given the diffusive variable \(X(t); t = 1, 2, \dots\), 
The Diffusion Entropy (DE) is defined as the Shannon entropy of the diffusion process $X(t)$ and has been widely applied to detect scaling properties in complex time series \cite{akin_jsmte09,grigolini2001asymmetric}.  % akin_pa06, 
The DE algorithm involves the following steps:
\begin{enumerate}
    \item Time Series Segmentation: for a given time lag $\Delta t$, divide the time series $X(t)$ into overlapping segments of duration $\Delta t$. Then calculate:
    $\Delta X (t, \Delta t) = X(t+\Delta t) - X(t),\ \forall\ t \in [0,t-\Delta t]$.
    \item Distribution Evaluation: for each time lag \(\Delta t\), determine the distribution \(p(\Delta x,\Delta t)\).
    \item Shannon Entropy Calculation: compute the Shannon entropy using the formula:
    \beq
    S(\Delta t) = - \int_{- \infty}^{+ \infty} p(\Delta x,\Delta t) \log p(\Delta x,\Delta t) dx
    \label{de_shannon}
    \eeq
    If the probability density function (PDF) is self-similar, i.e., \(p(\Delta x,\Delta t) = f(\Delta x / \Delta t^\delta)/\Delta t^\delta\), then:
    \beq
    S(\Delta t) = A + \delta \log (\Delta t + T)
    \label{de_scaling}
    \eeq
    To verify the accuracy of Eq. \eqref{de_scaling} on the data and to estimate the exponent $\delta$, it is sufficient to perform a linear best fit on a logarithmic scale along the time axis.
\end{enumerate}

%
% Acknowledgements and Disclosure of Interests
%
\begin{credits}
\subsubsection{\ackname} This work was supported by the Next-Generation-EU programme under the funding schemes PNRR-PE-AI scheme (M4C2, investment 1.3, line on AI)
FAIR “Future Artificial Intelligence Research”, grant id PE00000013, Spoke-8: Pervasive AI.

\subsubsection{\discintname} The authors have no competing interests to declare that are relevant to the content of this article.
\end{credits}

%
% References
%
\bibliographystyle{splncs04}
\bibliography{References}

\end{document}